\documentclass[preprint,doublespacing]{elsart}
\usepackage{txfonts}
\usepackage{cases}
\usepackage{graphicx}
\usepackage{longtable}

\journal{Nuclear Physics A}

\begin{document}

\begin{frontmatter}

\title{Theoretical study on nuclear structure by the multiple Coulomb scattering and magnetic scattering of relativistic electrons}
\author[1]{Jian Liu \corauthref{cor}}\corauth[cor]
{Corresponding author} \ead{liujian@upc.edu.cn.}
\author[2]{Xin Zhang}
\author[2]{Chang Xu}
\author[2]{Zhongzhou Ren}
\address[1]{College of Science,  China University of Petroleum (East China),
 Qingdao 266580, China}
\address[2]{Department of Physics, Nanjing University, Nanjing 210093, China}

\begin{abstract}

Electron scattering is an effective method to study the  nuclear structure. For the odd-$A$ nuclei with proton holes in the outmost orbits,  we investigate the contributions of proton holes to the nuclear quadrupole moments $Q$ and magnetic moments $\mu$ by the multiple Coulomb scattering and magnetic scattering. The deformed nuclear charge densities are constructed by the relativistic mean-field (RMF) models. Comparing the theoretical Coulomb and magnetic form factors with the experimental data, the nuclear quadrupole moments $Q$ and nuclear magnetic moments $\mu$ are investigated. From the electron scattering, the wave functions of the proton holes of odd-$A$ nuclei can be tested, which can also reflect the validity of the nuclear structure model.

\end{abstract}
\end{frontmatter}

\maketitle
\section{Introduction} \label{sec.1}

Electron scattering at high energies is one of the most powerful tools to investigate the nuclear structure, because the electron-nucleus interaction is well understood \cite{Hofstadter1956,Herbert,Guerra,Sick2014,Donnelly2015,Sick2015}.  According to the contributions from the Coulomb and magnetic interactions, electron scattering can be divided into the Coulomb scattering (or longitudinal scattering) and magnetic scattering \cite{Willey1963}. In the past decades, the nuclear charge densities of many stable have been determined accurately by the electron Coulomb scattering \cite{Vries1987,Angeli2013}.
Recently with the development of Radioactive Ion Beam (RIB) facilities, some exotic nuclei  far away from the $\beta$ stability line  with short half-lives can be produced \cite{Tanihata1995,Mueller2001}.  It is valuable for us to explore the charge density distributions of exotic nuclei with the electromagnetic probes. For this purpose, the new generation electron scattering facilities are under construction at RIKEN \cite{Wakasugi2008,Suda2009,Suda2011} and GSI \cite{GSI2006,GSI2004}. It is expected that in the near future elastic electron scattering off exotic nuclei can be achieved \cite{Suda2005,Simon2007}.

Besides the nuclear Coulomb scattering, the electron magnetic  scattering is another fundamental method to explore the nuclear magnetic properties and valence nucleon orbits \cite{Donnelly1984}. Different from the Coulomb scattering where all the charged nucleons contribute equally, the magnetic electron scattering is largely due to contributions of the unpair nucleon in the outmost orbit. The electron magnetic scattering provides a model-independent method to determine  the nuclear shell energy level and wave functions of the valence nucleon. Therefore the electron magnetic scattering experiments have been widely used to investigate the valence structure of the nuclei \cite{Lapikas1973,York1979,Euteneuer1977,Hicks1982,Baghaei1990,Miessen1984,Kalantar1988}.

With the rapid development of the experiments, great efforts have also been devoted to the theoretical studies of electron scattering off nuclei in the last decade \cite{Antonov2005,Cooper2005,Wang2006,Karataglidis2007,Dong2007,Roca2008,Chu2009,Gosselin2009,Liu2011,Liu2013,Roca2013,Jassim2014,Wang2015}.
The plane-wave Born approximation (PWBA) is a simple method to solve the electron scattering where the effects of nuclear Coulomb field on the scattering electrons are neglected. The phase-shift analysis method is an effect way to study the electron scattering which includes the Coulomb distortion effects by the exact phase-shift analysis of Dirac equations. This calculation method is also referred to as the distorted-wave Born approximation (DWBA) \cite{Yennie1954,Amundsen2014}. The phase-shift analysis method is an accurate method to calculate the cross sections of scattering electrons and has been applied in many theoretical studies \cite{Roca2008,Chu2009,Gosselin2009,Liu2011,Liu2013,Roca2013,Jassim2014}.

During the studies of electron scattering, the nuclear form factors  can be decomposed into several multipoles according to the selection rules, such as $C_0$, $C_2$ and $M_1$. For the Coulomb scattering off spherical nuclei, only the  $C_0$  contributions need to be taken into consideration. For the deformed nuclei, the high multipoles  $C_\lambda (\lambda\geq2)$ can reflect the contributions of nuclear deformation.
In the last decade, some researches focused on the Coulomb scattering off spherical nuclei where the $C_0$ contributions are precisely calculated under the DWBA method \cite{Wang2006,Roca2008,Chu2009,Roca2013}. As we know, most of the nuclei are found to be deformed in their ground states both theoretically and experimentally \cite{Moller1995,Raman2001,Scamps2013}, and the high multipoles  $C_\lambda (\lambda\geq2)$ need bo be included during the studies.  In Ref. \cite{Horowitz2014},  the dependence of the cross section on the parameter $\xi$ of DWBA calculations of $C_2$ multipole were investigated for $^{27}$Al. For the $C_0$ and $C_2$ contributions, there are aslo detailed comparative analysis on the PWBA and DWBA calculations in Refs. \cite{Moreno2009,Guerra2010,Nishimura1983,Nishimura1985}.

On the basis of studies done before, in this paper the  Coulomb  and magnetic multipoles are systematically investigated for the odd-$A$ nuclei with proton holes in the outermost orbits. The nuclei $^{14}$N, $^{27}$Al and $^{39}$K are chosen as the candidate nuclei. Their nuclear quadrupole and magnetic moments are assumed as the contributions of the proton holes in the outermost orbits. The density distributions of deformed nuclei are constructed by the relativistic mean-field (RMF) model.  After providing the deformed density distribution, the nuclear longitudinal form factors $F_L(q)$ are derived under the PWBA method. Then we take into account the Coulomb distortion corrections. By comparing the theoretical $F_L(q)$ with the experimental data, the nuclear quadrupole moments $Q$ can be extracted from the electron scattering experiments.

Besides the nuclear longitudinal scattering, the nuclear magnetic form factors $F_M(q)$  are also investigated in this paper. By the longitudinal scattering, we can analyze the contributions of proton hole to the total charge density distributions. However, by the magnetic scattering, we can further test the wave functions and energy levels of the proton hole. The wave functions of the proton hole are calculated under the theoretical framework of RMF model. Then we investigate the nuclear magnetic moments $\mu$ and compare them with the experimental data. Besides, the nuclear magnetic form factors $F_M(q)$ are further studied where the magnetic multipoles $M_L$ are derived under the PWBA method. The results are also compared with the experimental data. Combining the Coulomb scattering and magnetic scattering, the nuclear quadrupole moments $Q$ and magnetic moments $\mu$ can be extracted, which tests the validity of the nuclear structure model.

The paper is organized as follows. In Sec. \ref{sec.21}, the deformed density distribution model is constructed and in Sec. \ref{sec.22}, the theoretical framework of Coulomb and magnetic scattering is presented. In Sec. \ref{sec.3}, the nuclear longitudinal form factors $|F_{L}(q)|^2$ are calculated and the nuclear quadrupole moments $Q$ are extracted. In Sec. \ref{sec.4}, we study the nuclear magnetic moments $\mu$ and nuclear magnetic form factors $|F_{M}(q)|^2$.  Finally, a summary is given in Sec. \ref{sec.5}.

\section{Formalism}
\label{sec.2}

In this section, the main formulas of electron-nucleus scattering are outlined. The calculation scheme is divided into two parts. First, the nuclear density distribution is constructed where the nuclear deformation and magnetic moment are assumed as the contributions of proton hole in the outermost orbit. Then according to the deformed density model, the nuclear Coulomb form factors $|F_{L}(q)|^2$ and magnetic form factors $|F_{M}(q)|^2$ are investigated.

\subsection{Deformed density distribution model}
\label{sec.21}

In order to study the nuclear longitudinal form factors, the nuclear proton density distribution is constructed as the superposition of a spherical part and deformed part:
\begin{eqnarray}\label{nu_decom}
\rho(\mathbf{r})&=&\rho^0_p(r)+\rho^d_p(\mathbf{r})
\end{eqnarray}
where $\rho^0(r)$ is spherical symmetric and $\rho^d_p(\mathbf{r})$ describes the nuclear deformation. When the nuclear deformation is not big, $\rho^0_p(r)$ is considered as the main part of $\rho_p(\mathbf{r})$ and $\rho^d_p(\mathbf{r})$ is referred to as the deformed correction to $\rho^0_p(r)$.

For the nuclei with a proton hole in a closed shell, the nuclear deformation is assumed as the contributions of the proton hole. The relativistic mean-field (RMF) model is used to calculate the deformed proton density distribution. The advantage of RMF model is that the spin degrees of freedom of nucleons are treated microscopically and the spin-orbit splitting is given automatically, since it is essentially a relativistic
effect. In RMF model, the spherical symmetric part $\rho^0_p(r)$ can be calculated by \cite{Ring1996}:
\begin{eqnarray}\label{rho0}
\rho^0_p(r)=\sum_{i=1}^A(\frac{1}{2}-t_i)(|G_i(r)|^2+|F_i(r)|^2),
\end{eqnarray}
where $G_i$ and $F_i$ are the upper and lower components of the Dirac spinors for the occupied states. For the deformed part $\rho^d_p(\mathbf{r})$, the contribution of the proton hole can be described as \cite{Horowitz2014}:
\begin{eqnarray}\label{decom_pro_1}
\rho^d_p(\mathbf{r})=\beta_d\frac{1}{r^2}(|G_j(r)|^2+|F_j(r)|^2)(\frac{1}{4\pi}-|Y_{lm}(\theta,\varphi)|^2),
\end{eqnarray}
where $l$ and $j$ are the orbital and total angular momentum of the proton hole, respectively. The total proton number $Z$ is not changed because $\int\rho^d_p(\mathbf{r})\mathrm{d}^{\,3}r=0$. The parameter $\beta_d$ is introduced to adjust the nuclear quadrupole moment:
\begin{eqnarray}\label{qua}
Q=\frac{1}{e}\int_V\rho_p(\mathbf{r})(3z^2-r^2)\mathrm{d}\tau.
\end{eqnarray}
The spherical harmonics $|Y_{lm}(\theta,\varphi)|^2$ in Eq. (\ref{decom_pro_1}) can be further expanded:
\begin{eqnarray}\label{spherical_expand}
|Y_{lm}(\theta,\varphi)|^2
=(-1)^m\sum_L\left\{\frac{(2l+1)^2}{4\pi(2L+1)}\right\}^{\frac{1}{2}}\langle l\,m,l-m|L0\rangle \langle l0,l0|L0\rangle Y_{L0},
\end{eqnarray}
where $\langle l_1m_1, l_2m_2|LM\rangle$ is the Clebsch-Gordan (C. G.) coefficient. Substituting Eqs. (\ref{rho0}), (\ref{decom_pro_1}) and (\ref{spherical_expand}) into Eq. (\ref{nu_decom}), the deformed proton density distribution has the following form:
\begin{eqnarray}\label{2pF_decom_pro_2}
\rho_p(\mathbf{r})=\rho^0_p(r)+\sum_{L=2}^{2l}\rho^L_p(r)Y_{L0},
\end{eqnarray}
where
\begin{eqnarray}\label{rho_l}
\rho^L_p(r)=(-1)^m\beta_d\left\{\frac{(2l+1)^2}{4\pi(2L+1)}\right\}^{\frac{1}{2}}\langle l\,m,l-m|L0\rangle \langle l0,l0|L0\rangle (|G_j(r)|^2+|F_j(r)|^2). \nonumber \\
\end{eqnarray}
For the proton hole with the orbital angular momentum $l$, the spherical harmonics expanded value $L$ in Eq. (\ref{rho_l}) can take all the even numbers from $2$ to $2l$ according to the properties of C. G. coefficients. With the Eqs. (\ref{rho0}) and (\ref{rho_l}), the deformed proton density distribution in Eq. (\ref{2pF_decom_pro_2}) can be calculated.

\subsection{Nuclear electromagnetic form factors}
\label{sec.22}

During the scattering process, the electron scattering cross section can be calculated by the formula:
\begin{eqnarray}\label{cross}
\frac{\mathrm{d}\sigma}{\mathrm{d}\Omega}=\left(\frac{\mathrm{d}\sigma}{\mathrm{d}\Omega}\right)_M\cdot\eta\cdot|F(q)|^2,
\end{eqnarray}
with the Mott cross section
\begin{eqnarray}\label{Mott}
\left(\frac{\mathrm{d}\sigma}{\mathrm{d}\Omega}\right)_M=\frac{Z^2\alpha^2\,\mathrm{Cos}^2\frac{\theta}{2}}{4E_i^2\,\mathrm{Sin}^4\frac{\theta}{2}},
\end{eqnarray}
and the recoil factor $\eta=[1+\frac{2E}{M}\mathrm{Sin}^2(\theta/2)]^2$.
The form factor $|F(q)|^2$ can be further developed as:
\begin{eqnarray}\label{Formfactor}
|F(q)|^2=|F_L(q)|^2+(1+2\,\mathrm{tan}^2\frac{\theta}{2})|F_M(q)|^2.
\end{eqnarray}
$|F_L(q)|^2$ is the longitudinal term (or Coulomb term) and $|F_M(q)|^2$ is the magnetic term. The longitudinal and magnetic form factor can be decomposed into several multipoles according to the selection rules:
\begin{eqnarray}\label{Clambda}
|F_L(q)|^2=\sum_{\lambda=0,\,\mathrm{even}}^{\infty}|F_{C\lambda}(q)|^2, \ \mathrm{and} \ \   |F_M(q)|^2=\sum_{\lambda=1,\,\mathrm{odd}}^{\infty}|F_{M\lambda}(q)|^2.
\end{eqnarray}

\subsubsection{Longitudinal form factor}

The longitudinal term $F_L(q)$ in Eq. (\ref{Formfactor}) can be obtained by folding the proton form factor $F_p(q)$ with the electric form factor of a single proton $G_E^p(q)$:
\begin{eqnarray}\label{FL}
|F_L(q)|^2=G_E^p(q)^2\cdot|F_p(q)|^2,
\end{eqnarray}
where the detail form of $G_E^p(q)$ can be found in Refs. \cite{Kelly2002} and \cite{Horowitz2012}.

Under the PWBA method, the proton form factor $F_p(q)$ can be calculated by:
\begin{eqnarray}\label{FpPWBA}
F_p(q)=\frac{1}{Z}\int \mathrm{d}^3r \rho_p(\mathbf{r})e^{i\mathbf{q}\cdot\mathbf{r}},
\end{eqnarray}
where $\rho_p(\mathbf{r})$ is the deformed proton density of Eq. (\ref{2pF_decom_pro_2}). The exponential function $e^{i\mathbf{q}\cdot\mathbf{r}}$ in Eq. (\ref{FpPWBA}) can be expanded as a sum of the spherical harmonics:
\begin{eqnarray}\label{bessel_expand}
e^{i\mathbf{q}\cdot\mathbf{r}}=\sum_{L'}\sqrt{4\pi(2L'+1)}\cdot i^{L'}\cdot j_{L'}(qr)\cdot Y_{L'0}(\theta,\varphi),
\end{eqnarray}
where $j_{L}(qr)$ is the spherical Bessel function. Substituting Eqs. (\ref{2pF_decom_pro_2}) and (\ref{bessel_expand}) into Eq. (\ref{FpPWBA}) and taking into account the orthogonality and normalization of the spherical harmonics,
the proton form factor can be obtained by squaring $|F_p(q)|^2$ and averaging it over the orbital angular momentum projection $m$ from $-l$ to $l$. Through the derivation under the PWBA method, the proton form factor can also be divided into two parts:
\begin{eqnarray}\label{Fp2}
|F_p^{PW}(q)|^2=|F_{C0}^{PW}(q)|^2+|F_{C\lambda}^{PW}(q)|^2.
\end{eqnarray}
$|F_{C0}^{PW}(q)|^2$ in Eq. (\ref{Fp2}) can be seen as the contribution of the spherical part $\rho^0_p(r)$ on nuclear longitudinal form factor:
\begin{eqnarray}\label{sphe_PW}
|F_{C0}^{PW}(q)|^2=G_E^p(q)^2\cdot\left(\frac{1}{Z}\int\mathrm{d}^3r\,\rho^0_p(r)j_0(qr)\right)^2,
\end{eqnarray}
and $|F_{C\lambda}^{PW}(q)|^2$ can be seen as the contribution of the deformed part $\rho^d_p(\mathbf{r})$:
\begin{eqnarray}\label{defo_PW}
|F_{C\lambda}^{PW}(q)|^2=G_E^p(q)^2\cdot\frac{1}{2l+1}\sum_{m=-l}^{l}\left( \sum_{L=2,\,\mathrm{even}}^{2l}\langle l\,m,l-m|L0\rangle \langle l0,l0|L0\rangle I_L(q)\right)^2,
\end{eqnarray}
where
\begin{eqnarray}\label{IL}
I_L(q)=\frac{1}{Z}\int\mathrm{d}r\,j_L(qr)\beta_d(|G_j(r)|^2+|F_j(r)|^2).
\end{eqnarray}
For proton hole with the angular momentum $l$, there are only contributions for $L=0,2,4,\cdots,2l$, according to the symmetry property of C. G. coefficients and the orthogonality property of spherical harmonics.

The Coulomb distortion corrections are very significant during the theoretical studies of electron scattering \cite{Amundsen2014}.  Under the theoretical framework of Eq. (\ref{Fp2}) of PWBA, the longitudinal term $|F_L(q)|^2$ can be divided into two parts $|F_{C0}(q)|^2$ and $|F_{C\lambda}(q)|^2$. The Coulomb distortion corrections need to be included in these two parts and the  Eq. (\ref{Fp2}) can be rewritten as:
\begin{eqnarray}\label{FcDW2}
|F_L^{DW}(q)|^2&=&|F_{C0}^{DW}(q)|^2+|F_{C\lambda}^{DW}(q)|^2
\end{eqnarray}
The DWBA calculation $|F_{C0}^{DW}(q)|^2$ of $C_0$ multiple can be done by many methods, such as the relativistic Eikonal approximation \cite{Wang2006} and the phase-shift analysis method \cite{Chu2009,Liu2011,Liu2013}. In this paper, we calculate the $|F_{C0}^{DW}(q)|^2$ by the phase-shift analysis method and the detailed formulas can be seen in Refs. \cite{Chu2009,Yennie1954}.

For $C_\lambda$ multiples, there are few researches on the DWBA calculations $|F_{C\lambda}^{DW}(q)|^2$. In Refs. \cite{Nishimura1983,Nishimura1985}, the authors presented full DWBA calculations for the $C_2$ multipole of the inelastic electron scattering. It can be seen in the Fig. 1 of Ref. \cite{Nishimura1985} that the deviations between the cross sections $\sigma_{C2}^{DW}$ and  $\sigma_{C2}^{PW}$ mainly located at the diffraction minima. However at these positions, the nuclear longitudinal form factors $|F_L(q)|^2$ are mainly determined by the contributions of $C_0$ multipole, which can be seen in the left panels of Figs. \ref{fig:N14FSU}, \ref{fig:Al27FSU} and \ref{fig:K39FSU} of this paper. Besides,  when construct the deformed density distribution, we assume that for the small deformation, the spherical part $\rho^0_p(r)$ is the major part of $\rho_p(\mathbf{r})$ and the nonspherical part $\rho^d_p(\mathbf{r})$ is the high order correction of $\rho^0_p(r)$. Therefore, we calculate the spherical contributions $|F_{C0}(q)|^2$ accurately where the Coulomb distortion corrections are taken into account with the DWBA method. For $\rho^d_p(\mathbf{r})$, their contribution $|F_{C\lambda}(q)|^2$ is calculated with Eq. (\ref{defo_PW}) under the PWBA method. Finally, the longitudinal form factors of deformed nuclei is given as following:
\begin{eqnarray}\label{FcDW}
|F_L^{DW}(q)|^2\approx|F_{C0}^{DW}(q)|^2+|F_{C\lambda}^{PW}(q)|^2.
\end{eqnarray}
Combining the DWBA calculations of $|F_{C0}^{DW}(q)|^2$ and the formula of Eq. (\ref{defo_PW}), the nuclear longitudinal form factor $|F_L(q)|^2$ can be obtained.

\subsubsection{Magnetic form factor}

For odd-$A$  nuclei with the proton-hole configuration in a closed shell, the nuclear magnetic moments are assumed as the contributions of the proton hole. The wave function of proton hole can be described by the Dirac spinor under the RMF model:
\begin{eqnarray}\label{DiracSpinor}
\psi_{n\kappa m} = \left[ \begin{array}{ccc}
 i[G(r)/r]\Phi_{\kappa m} \\
 -[F(r)/r]\Phi_{-\kappa m} \end{array}\right]=\left[ \begin{array}{ccc}
 i|n\kappa m\rangle\\
 -\overline{|n\kappa m\rangle}\end{array}\right].
\end{eqnarray}

The magnetic form factor is defined as \cite{Forest1966,Donnelly1984}:
\begin{eqnarray}\label{magform}
F_M^2(q^2)=\frac{4\pi f_{\mathrm{sn}}^{\,2}(q)f_\mathrm{c.m.}^{\,2}(q)}{2j+1}\sum_{L=1,\mathrm{odd}}^{2j}\left|\langle j||\hat{T}_L^{\mathrm{mag}}||j\rangle\right|^2
\end{eqnarray}
where $j$ is the total angular momentum of the nucleus and the sum of magnetic multipole operator $ \hat{T}_L^{\mathrm{mag}} $ is done for all odd multipoles $L$ up to $2j$. $f_{\mathrm{sn}}(q)=[1+(q/855\mathrm{MeV})^2]^{-2}$ and $f_\mathrm{c.m.}(q)=\exp(q^2b^2/4A)$
are the single-nucleon factor and center-of-mass factor, respectively \cite{Dong2007}. In PWBA method, the magnetic multipole operator can be given as \cite{Blunden1991}:
\begin{eqnarray}
\hat{T}_L^{\mathrm{mag}}&=&\int d^3{r}j_L(qr){\mathbf{Y}} _{LL1}^M(\hat{r})\cdot\hat{{J}}(\mathbf{r})\nonumber\\
&=&\frac{i}{\sqrt{4\pi}}\int_0^\infty j_L(qr)\hat{J}_{LL}(r)r^2dr,
\label{Tmag}
\end{eqnarray}
where  ${\mathbf{Y}}_{LL1}^M(\hat{r})$ is vector spherical harmonics and $\hat{J}_{LL}(r)$ is the transition current
density operator. With the effective current operator and the four-component Dirac wave function, the multipole operator $\hat{T}_L^{\mathrm{mag}}$ can be written in a block matrix form \cite{Kim1986,Serot1981,Serot1986}:
\begin{eqnarray}\label{tmatrix}
\hat{T}_L^{\mathrm{mag}}=\left(
  \begin{array}{cc}
  iq(\lambda'/2M) \Sigma_L^{'M}(\mathbf{r}) &  Q\Sigma_L^{M}(\mathbf{r})\\
    Q\Sigma_L^{M}(\mathbf{r}) &  -iq(\lambda'/2M) \Sigma_L^{'M}(\mathbf{r})\\
  \end{array}
\right),
\end{eqnarray}
where
\begin{eqnarray}
\Sigma_L^{M}(\mathbf{r})&\equiv &\mathbf{M}_{LL}^M(\mathbf{r})\cdot \sigma,\nonumber \\
\Sigma_L^{'M}(\mathbf{r})&\equiv & -i(\nabla\times\mathbf{M}_{LL}^M(\mathbf{r}))\cdot\sigma/q,\nonumber\\
\mathbf{M}_{LL}^M(\mathbf{r})&\equiv &j_L(qr) \mathbf{Y}_{LL}^M(\hat{r}).\nonumber
\end{eqnarray}
The single-particle reduced matrix elements $\langle j||\hat{T}_L^{\mathrm{mag}}||j\rangle$ can be derived by using the formulas in Refs. \cite{Willey1963,Bjorken1964,Suhonen2007}. Then substituting Eqs.  (\ref{Tmag}) and  (\ref{tmatrix}) into Eq. (\ref{magform}), the magnetic form factor can be reduced to the following:
\begin{eqnarray}\label{magform2}
|F_M(q)|^2=f_{\mathrm{sn}}^{\,2}(q)f_\mathrm{c.m.}^{\,2}(q)\sum_{L=1,\mathrm{odd}}^{2j}|F_{ML}(q)|^2,
\end{eqnarray}
where the $L$th magnetic multipole $F_{ML}(q)$ is the Fourier transform of the transition current density $J_{LL}$:
\begin{eqnarray}\label{FML}
F_{ML}(q)=\frac{1}{\sqrt{2j+1}}\int_0^\infty J_{LL}(r)j_L(qr)r^2dr.
\end{eqnarray}
The current density $J_{LL}(r)$ in Eq. (\ref{FML}) can be divided into two parts:
\begin{eqnarray}\label{JLL}
J_{LL}(r)\equiv \langle j||\hat{J}_{LL}(r)||j\rangle=J_{LL}^c(r)+J_{LL}^s(r).
\end{eqnarray}
The convection part $J_{LL}^c(r)$ is due to the orbital motion of proton:
\begin{eqnarray}\label{JC}
J_{LL}^c(r)&=&2Q{\left( { - 1} \right)^{{l}}}\left( {2j + 1} \right)\sqrt {\left( {2L + 1} \right)}\nonumber\\
&& \times\left( {\begin{array}{ccc}
{j}&{j}&L\\
{1/2}&{ 1/2}&-1
\end{array}} \right){G}\left( r \right){F}\left( r \right)/{r^2},
\end{eqnarray}
and the spin part $J_{LL}^s(r)$ is due to the spin of nucleon:
\begin{eqnarray}\label{JS}
J_{LL}^s(r)&=& - \frac{{\kappa}}{{2{M}}}{\left( { - 1} \right)^{j + \frac{1}{2} }}\left( {2j + 1}
 \right)\sqrt {L\left( {L + 1} \right)\left( {2L + 1} \right)} \nonumber\\
 &&\times\left( {\begin{array}{*{20}{c}}
{j}&{j}&L\nonumber\\
{1/2}&{-1/2}&0
\end{array}} \right)\left[ {{G^2}\left( r \right) - {F^2}\left( r \right)} \right]/{r^3}\\
&& + \frac{{\kappa}}{{2{M}}}{\left( { - 1} \right)^l}\left( {2j + 1} \right)
\sqrt {\left( {2L + 1} \right)} \left( {\begin{array}{*{20}{c}}
{j}&{j}&L\\
{1/2}&{1/2}&-1
\end{array}} \right)\nonumber\\
&&\times\frac{1}{r^2}\left( {\frac{d}{{dr}} - \frac{1}{r}} \right)\left[ {{G^2}\left( {r} \right)
 + {F^2}\left( r \right))} \right],
\end{eqnarray}

where $Q$ and $\kappa$ are the charge and the anomalous magnetic moment of the nucleon, respectively. Substituting Eqs. (\ref{FML})-(\ref{JS}) into Eq. (\ref{magform2}), the magnetic form factor can be obtained in PWBA method.

Besides, with the current density $J_{LL}(r)$ the magnetic moments of odd-$A$ nuclei can also be calculated from the $q\rightarrow 0$ limit of the form factor \cite{Blunden1991}:
\begin{eqnarray}\label{Mmoments}
\mu=2M\sqrt{\frac{j}{6(j+1)(2j+1)}}\int_0^\infty dr\,r^3\,J_{11}(r).
\end{eqnarray}

\section{Longitudinal scattering}
\label{sec.3}

In this section, the nuclear longitudinal form factors are investigated with the formulas presented in Sec. II. The nuclei $^{14}$N, $^{27}$Al and  $^{39}$K are chosen as the candidate nuclei, where the nuclear deformation of these nuclei are considered as the contributions of proton holes in the outermost orbits.

The wave functions of nucleons and proton holes are obtained under the RMF theory. In order to test the validity of RMF theory, we firstly investigate the ground state properties of these nuclei and compare them with the experimental data. The binding energies per nucleon $B/A$ and charge radii $R_C$ of $^{14}$N, $^{27}$Al and $^{39}$K are calculated with the FSU and NL3* parameter set, respectively, and the results are presented in Table \ref{table1}. In this table, one can see that the theoretical results of $B/A$ and $R_C$ calculated from the RMF theory are in agreement with the experimental data. This suggests that the RMF theory, can reproduce the ground properties of these nuclei.

\subsection{$F_L(q)$ of $^{14}$N}

In RMF model, the configuration of protonic arrangement of $^{14}$N is $(1s_{1/2})^2(1p_{3/2})^2(1p_{1/2})^1$. In the orbit $(1p_{1/2})^1$ of $^{14}$N, there exists a proton hole with the orbital angular momentum $l=1$. The wave functions $G_i$ and $F_i$ of protons in different energy levels can be obtained by solving the Dirac equations. After obtaining the wave functions of nucleons and proton hole, the spherical part of proton density $\rho_p^0(r)$ can be calculated with Eq. (\ref{rho0}). Besides, with the Eq. (\ref{rho_l}), the deformed part $\rho^d_p(\mathbf{r})$ of $^{14}$N can also be obtained.

In Fig. \ref{fig:density14N}, we display the theoretical proton density distribution $\rho_p(\mathbf{r})$ of $^{14}$N in cylindrical coordinates where the corresponding quadrupole moments are $Q=0.0\ \mathrm{fm}^2$ and $Q=3.2\ \mathrm{fm}^2$, respectively. For the case of $Q=0.0\ \mathrm{fm}^2$, the proton density distribution is spherical symmetric. When the deformed part $\rho^d_p(\mathbf{r})$ is included, there is a small deformation for $\rho_p(\mathbf{r})$, which can give the quadrupole moment $Q=3.2\ \mathrm{fm}^2$. After the proton density distribution $\rho_p(\mathbf{r})$ is obtained, the nuclear charge form factor can be further investigated with formulas in Sec. IIB.  For the nucleus $^{14}$N, the proton hole in $1p_{1/2}$ energy level has the orbital angular momentum $l=1$. Through the theoretical derivation  with Eq. (\ref{defo_PW}), the contribution of the deformed part $\rho^d_p(\mathbf{r})$ is:
\begin{eqnarray}\label{Fpw14N}
|F_{C2}^{PW}(q)|^2=2I_2(q)^2.
\end{eqnarray}
Substituting Eq. (\ref{Fpw14N}) into Eq. (\ref{FcDW}), the nuclear longitudinal form factor for $^{14}$N is:
\begin{eqnarray}\label{Fcw14N}
|F_L(q)|^2=|F_{C0}^{DW}(q)|^2+2I_2(q)^2.
\end{eqnarray}

The nuclear longitudinal form factors of $^{14}$N are calculated by Eqs. (\ref{Fpw14N}) and (\ref{Fcw14N}), and the theoretical results are displayed in the right panel of Fig. \ref{fig:N14FSU}. In this figure, one can see that at most momentum transfers, the spherical contributions $|F_{C0}^{DW}(q)|^2$ (the dot line $Q=0.0\ \mathrm{fm}^2$) are in good agreement with the experimental data. However at the diffraction minimum of the form factor, $|F_{C0}^{DW}(q)|^2$  is significantly smaller than the experimental data. By including the quadrupole contribution $|F_{C2}^{PW}(q)|^2$, the deviations between the theoretical results and experimental data gradually reduce. For the form factors at other positions, there is very little changes on the theoretical results with the introduction of quadrupole contribution $|F_{C2}^{PW}(q)|^2$. In order to interpret these results, we also present $C_0$, $C_2$ and total form factor in the left panel of Fig. \ref{fig:N14FSU}, which are calculated by Eqs. (\ref{sphe_PW}) and (\ref{defo_PW}) under the PWBA method. It can be seen in this figure that at most momentum transfers, the $C_2$ contribution is  much smaller than the $C_0$. Only in the diffraction minimum and high momentum transfers, the $C_2$ contribution plays a major role in the total form factors. Therefore at these positions, the nuclear longitudinal form factors are significantly modified by introducing the quadrupole contributions $C_2$.

\subsection{$F_L(q)$ of $^{27}$Al}

In this part, the longitudinal form factors of $^{27}$Al are investigated. As before, we first calculate the proton density distribution of $^{27}$Al. For the nucleus $^{27}$Al, the configuration of proton arrangement is $(1s_{1/2})^2(1p_{3/2})^2(1p_{1/2})^2(1d_{5/2})^5$, which has a proton hole in the orbit $1d_{5/2}$ with the orbital angular momentum $l=2$. The deformed proton density $\rho_p(\mathbf{r})$ and its spherical part $\rho_p^0(r)$ are calculated by Eqs. (\ref{rho0}), (\ref{2pF_decom_pro_2}) and (\ref{rho_l}) with the FSU parameter set.  Fig. \ref{fig:density27Al} presents the proton density distributions of $^{27}$Al where the corresponding quadrupole moments are $Q=0.0\ \mathrm{fm}^2$ and $Q=8.3\ \mathrm{fm}^2$, respectively. One can see that there exists a central depression in the left panel of Fig. \ref{fig:density27Al}, because the $2s_{1/2}$ state of $^{27}$Al is unoccupied and the $s$-state protons are lacked. With the nonspherical part $\rho^{d}_p(\mathbf{r})$ included, there are two regions of pronounced localization at the outer ends of symmetry axis and an oblate deformed core in the right panel of Fig. \ref{fig:density27Al}. This distribution is very close to the Fig. 3 of Ref. \cite{Ebran2013} where the self-consistent ground-state density of $^{28}$Si is calculated from the deformation-constrained RMF theory with DD-ME2 parameter set.  This indicates that the density model in Sec. IIA can provide a reasonable description of the deformed proton distribution, because $^{27}$Al has only one less proton than $^{28}$Si.

When the deformed proton density $\rho_p(\mathbf{r})$ of $^{27}$Al is obtained, we investigate the corresponding longitudinal form factor. For the nucleus $^{27}$Al, the proton hole in $1d_{5/2}$ energy level has the orbital angular momentum $l=2$. The contribution of the deformed part $\rho^{d}_p(\mathbf{r})$ to the longitudinal form factor can be calculated with Eq. (\ref{defo_PW}):
\begin{eqnarray}\label{Fpw27Al}
|F_{C\lambda}^{PW}(q)|^2=\frac{10}{7}I_2(q)^2.
\end{eqnarray}
Substituting Eq. (\ref{Fpw27Al}) into Eq. (\ref{FcDW}), the longitudinal form factor for $^{27}$Al is:
\begin{eqnarray}\label{Fcw27Al}
|F_L(q)|^2=|F_{C0}^{DW}(q)|^2+\frac{10}{7}I_2(q)^2.
\end{eqnarray}
The  $C_4$ multipole is small, therefore, in Eq. (\ref{Fcw27Al}) only the $C_2$  multipole is taken into considerations. Besides, in Ref. \cite{Horowitz2014} the author have calculated the longitudinal form factor of $^{27}$Al. In this paper, general formulas Eqs. (\ref{defo_PW}) and (\ref{FcDW}) for the nuclear longitudinal form factors are presented. Based on these two formulas, we have derived the Eq.  (\ref{Fcw27Al}) which is consistent with the researches of Ref. \cite{Horowitz2014}.

With Eq. (\ref{Fcw27Al}), the longitudinal form factors of $^{27}$Al are calculated and the results are presented in the right panel of Fig. \ref{fig:Al27FSU} where the corresponding quadrupole moments are $Q=0.0\ \mathrm{fm}^2$ and $Q=8.3\ \mathrm{fm}^2$, respectively. From this figure, one can also see that at the diffraction minima, the deviations between $|F_{C0}^{DW}(q)|^2$ (the dot lines $Q=0.0\ \mathrm{fm}^2$) and experimental data are evident. By investigating the longitudinal form factors with the deformed scattering model (the solid line $Q=8.3\ \mathrm{fm}^2$), the deviations at the diffraction minima become smaller. Besides, in Figs. \ref{fig:Al27FSU} one can also see that at higher momentum transfers ($q>2.5\ \mathrm{fm}^{-1}$), the results of spherical contribution $|F_{C0}^{DW}(q)|^2$  drops more quickly than the experimental data. With the quadrupole contribution $|F_{C2}^{PW}(q)|^2$ taken into account, the theoretical longitudinal form factor coincides with the experimental data much better at high momentum transfers. In order to explain the results of right panel of Fig. \ref{fig:Al27FSU}, we also present $C_0$ and $C_2$ in the left panel of Fig. \ref{fig:Al27FSU}, which are calculated under the PWBA method. One can see that at the diffraction minima and high momentum transfers, adding the $C_2$ multipole can significantly modify the total longitudinal form factors.

\subsection{$F_L(q)$ of $^{39}$K}

Besides $^{14}$N and $^{27}$Al, we also investigate the longitudinal form factors of $^{39}$K, which has a proton hole in $1d_{3/2}$ orbit with $l=2$. From Eqs. (\ref{rho0}), (\ref{2pF_decom_pro_2}) and (\ref{rho_l}), the deformed proton density $\rho_p(\mathbf{r})$ is calculated and presented in Fig. \ref{fig:density39K} where the corresponding quadrupole moments are $Q=0.0\ \mathrm{fm}^2$ and $Q=3.9\ \mathrm{fm}^2$, respectively.  It can be seen from this figure that with the nonspherical part $\rho^{d}_p(\mathbf{r})$ included, it also appears the localization for the proton distributions of  $^{39}$K, which is similar to the result of Fig. \ref{fig:density27Al}.

After the proton density $\rho_p(\mathbf{r})$ of $^{39}$K is obtained, the corresponding longitudinal form factor is calculated. The proton hole of of $^{39}$K is in $1d_{3/2}$ orbit with the angular momentum $l=2$. Substituting the deformed proton density $\rho_p(\mathbf{r})$ of $^{39}$K  into Eq. (\ref{Fp2}) we can obtain:
\begin{eqnarray}\label{Fpw39K}
|F_{C\lambda}^{PW}(q)|^2&=&C_2^2(q)+C_4^2(q)\nonumber\\
&=&\frac{10}{7}I_2(q)^2+\frac{18}{7}I_4(q)^2.
\end{eqnarray}
Then substituting Eq. (\ref{Fpw39K}) into Eq. (\ref{FcDW}), the longitudinal form factor for $^{39}$K is:
\begin{eqnarray}\label{Fcw39K}
|F_L(q)|^2=|F_{C0}^{DW}(q)|^2+\frac{10}{7}I_2(q)^2+\frac{18}{7}I_4(q)^2.
\end{eqnarray}

With Eq. (\ref{Fcw39K}), we calculate longitudinal form factors of $^{39}$K and the results are presented in the right panel of Fig. \ref{fig:K39FSU} where the corresponding quadrupole moments are $Q=0.0\ \mathrm{fm}^2$ and $Q=3.9\ \mathrm{fm}^2$, respectively. From this figure, it can also be seen that there are significant deviations between $|F_{C0}^{DW}(q)|^2$ (the dot line $Q=0.0\ \mathrm{fm}^2$) and experimental data. With the quadrupole contribution taken into account, the theoretical longitudinal form factor (the solid line $Q=3.9\ \mathrm{fm}^2$) coincides with the experimental data much better at high momentum transfers. In order to interpret this result, in the left panel of Fig. \ref{fig:K39FSU} we present $C_0$, $C_2$ and $C_4$ multipoles. It can be seen in this panel that the influence of $C_2$ is small at these range and the modification to $|F_L(q)|^2$ at high momentum transfers is due to contributions of $C_4^2(q)$.

Now we make a brief summary to Section III. From Figs. (\ref{fig:N14FSU}), (\ref{fig:Al27FSU}) and (\ref{fig:K39FSU}), one can see that with the electron quadrupole scattering taken into considerations, the theoretical longitudinal form factors $|F_L(q)|^2$ coincide with the experimental data much better. The quadrupole moments $Q$ extracted from the electron scattering are presented in Table \ref{table2}. For comparison, we also present the previous experimental $Q$ in Table \ref{table2}. It can be seen that the quadrupole moments $Q$ extracted from the electron scattering are in an acceptable range, which is smaller than twice the previous experimental value.

\section{Magnetic scattering}
\label{sec.4}

In this section, the nuclear magnetic moments and magnetic form factors of $^{27}$Al and $^{39}$K are investigated. Because the neutron number of these two nuclei are even, their nuclear magnetic moments $\mu$ can be seen as the contributions of the proton hole in the outmost shell. The nuclear magnetic form factor can be calculated with Eqs. (\ref{magform})-(\ref{JS}). However, because of the nuclear many-body effects, there are deviations between the theoretical magnetic form factors and experimental data, which can be seen in Figs. \ref{fig:Mag27Al} and \ref{fig:Mag39K}. In order to treat these discrepancies, the spectroscopic factors $\alpha_L$ are introduced to the magnetic multipoles and the Eq. (\ref{magform2}) can be rewritten as:
\begin{eqnarray}\label{magform3}
|F_M(q)|^2=f_{\mathrm{sn}}^{\,2}(q)f_\mathrm{c.m.}^{\,2}(q)\sum_{L=1,\mathrm{odd}}^{2j}\alpha_L^2F_{ML}^2(q),
\end{eqnarray}
where $F_{ML}(q)$ can be calculated by Eq. (\ref{FML}). In the low $q$ region, the $M1$ multipole dominates the electron scattering, therefore, the factor $\alpha_1$ is fixed to the ratio of the experimental magnetic moment to the single-particle value: $\alpha_1=\mu_\mathrm{exp}/\mu_\mathrm{sp}$. The other $\alpha_L$s are usually obtained by fitting the theoretical form factors to the experimental data \cite{Hicks1982,Baghaei1990}. In Table \ref{table3}, we present the theoretical nuclear magnetic moments $\mu_\mathrm{sp}$ of $^{27}$Al and $^{39}$K, which are calculated by Eq. (\ref{Mmoments}) under the RMF model with FSU and NL3* parameter sets, respectively. The experimental data are taken from the Refs. \cite{Stone2005,Raghavan1989}. From Table \ref{table3}, it can be seen that the single-particle nuclear magnetic moments $\mu_\mathrm{sp}$ coincide with the experimental values, which means single-particle levels of RMF model is reliable.

\subsection{$|F_M(q)|^2$ of $^{27}$Al}

After obtaining the nuclear magnetic moments $\mu$, the nuclear magnetic form factor $|F_M(q)|^2$ of $^{27}$Al is investigated. The proton hole of $^{27}$Al has the angular momentum $j=\frac{5}{2}$, which makes the magnetic multipoles $M1$, $M3$ and $M5$ need to be taken into accounts. In Fig. \ref{fig:Mag27Al}, we present the theoretical magnetic form factors of $^{27}$Al calculated with the FSU and NL3* parameter set, respectively. The dash-dot curves correspond to the results of the pure contribution of the $1d_{5/2}$ proton hole. The solid curves denote the results of Eq. (\ref{magform3}) which include the corrections of spectroscopic factors. In Table \ref{table4}, the spectroscopic factors for $^{27}$Al are presented where $\alpha_1$ is fixed to be $\mu_\mathrm{exp}/\mu_\mathrm{sp}$ and other $\alpha_L$s are obtained by the best fitting to the experimental data. One can see in Fig. \ref{fig:Mag27Al} that the pure contribution of the $1d_{5/2}$ proton hole can give a close description of the shape of experimental data. With the spectroscopic factors included in Eq. (\ref{magform2}), the theoretical results coincide with the experimental data very well. Therefore, it once again proves that there exists a proton hole in $1d_{5/2}$ level for $^{27}$Al.  Combining the magnetic moments $\mu$ of Table \ref{table3} and magnetic form factors $|F_M(q)|^2$ of Fig. \ref{fig:Mag27Al}, the Dirac spinor of the proton hole calculated under the RMF model can be tested and the information of the nuclear energy level can be studied.

\subsection{$|F_M(q)|^2$ of $^{39}$K}

Besides $^{27}$Al, the  magnetic form factors $|F_M(q)|^2$ of $^{39}$K are also studied. In $1d_{3/2}$ orbit of $^{39}$K, there also exists a proton hole with the angular momentum $j=\frac{3}{2}$. From Eq. (\ref{magform3}), it can be seen that there are contributions from the $M1$ and $M3$ multipoles to the total $|F_M(q)|^2$. In Fig. \ref{fig:Mag39K}, we present the magnetic form factor $|F_M(q)|^2$ of $^{39}$K where the FSU and NL3* parameter sets are used. The dash-dot curves represent the results calculated by the Eq. (\ref{magform2}) where the pure contribution of the $1d_{3/2}$ proton hole is taken into account. The solid curves represent the results calculated from the Eq. (\ref{magform3}), which include the corrections of spectroscopic factors of Table \ref{table4}. Similar to the results of Fig. \ref{fig:Mag27Al}, Eq. (\ref{magform2}) (dash-dot curves in Fig. \ref{fig:Mag39K}) gives a close description of experimental magnetic form factor, which means the wave functions and energy levels obtained by the RMF model are reasonable. By including the spectroscopic factors further, Eq. (\ref{magform3}) (solid curves in Fig. \ref{fig:Mag39K}) can reproduce the experimental data better. Combining Table \ref{table3}, Figs. \ref{fig:Mag27Al} and \ref{fig:Mag39K}, the Dirac spinor of the proton holes of RMF model can be tested by comparing the theoretical magnetic moments and magnetic form factors with the experimental data.

\section{Summary}
\label{sec.5}

In this paper, we systematically study the nuclear structure of odd-A nuclei by the multiple Coulomb scattering and magnetic scattering of relativistic electrons. During the studies, the nuclear deformation and magnetic moments are considered as the contribution of the proton hole in the outermost orbit. The deformed nuclear density distributions $\rho_p(\mathbf{r})$ are constructed as the superposition of the spherical part $\rho_p^0(r)$ and deformed part $\rho^{d}_p(\mathbf{r})$. Both of the two parts are calculated under the RMF model. For the nuclei with smaller deformation, we assume the spherical symmetric part $\rho_p^0(r)$ is the major part,  and the nonspherical part $\rho^{d}_p(\mathbf{r})$ is the high order correction to $\rho_p^0(r)$.

After the density distributions are obtained, the nuclear proton form factors are investigated by the combination of the DWBA method and PWBA method. For the spherical contribution $C_0$, the Coulomb distortion corrections are taken into account by the accurate DWBA method. For the nonspherical contributions $C_\lambda$, we calculate under the PWBA method. Finally, the nuclear longitudinal form factors $|F_L(q)|^2$ is obtained by combining the spherical and quadrupole contributions. The theoretical $|F_L(q)|^2$ of nuclei $^{14}$N, $^{27}$Al and $^{39}$K are calculated and compared with the experimental data. By this way, the nuclear quadrupole moments $Q$ are extracted from the electron-nucleus scattering. By comparing with the experimental quadrupole moments $Q$ measured in previous experiments, one can see that the quadrupole moments $Q$  extracted from the electron scattering experiments in this paper are reasonable and consistent with the previous experimental value in an acceptable range.

Besides the nuclear longitudinal form factors $|F_L(q)|^2$, the nuclear magnetic form factors $|F_M(q)|^2$ are also investigated for the nuclei $^{27}$Al and $^{39}$K. Their nuclear magnetic moments $\mu$ are also considered as the contributions from the proton hole in the outmost orbit. By the electron magnetic scattering, the Dirac spinor of the proton hole can be tested because the nuclear magnetic form factors $|F_M(q)|^2$ are very sensitive to the wave function and angular momentum of the proton hole. The theoretical nuclear magnetic moments $\mu$ are also calculated under the RMF model and compared with the experimental data. Combining the studies of nuclear magnetic moments $\mu$ and nuclear magnetic form factors $|F_M(q)|^2$, the Dirac spinor and valence energy orbit of the proton hole can be investigated.
From the studies in this paper, one can see that the nuclear structure can be explored by the multiple Coulomb scattering and magnetic scattering of relativistic electrons, which also provide a guide for the future electron scattering experiments of exotic nuclei.

\begin{center}
{\large Acknowledgments }
\end{center}
This work is supported by the National Natural Science
Foundation of China (Grants Nos. 11505292, 11175085, 11235001, and 11447226), by the Shandong Provincial Natural Science Foundation, China (Grant No. BS2014SF007), by the Fundamental Research Funds for the Central Universities  (Grants
Nos. 15CX02072A, 15CX02070A, 15CX05026A, 13CX10022A, and 14CX02157A).

\newpage
Table Captions:

Table.~\ref{table1} Binding energies per nucleon $B/A$ (MeV) and charge radii $R_C$ (fm) of nuclei $^{14}$N, $^{27}$Al and $^{39}$K, calculated under the RMF theory  with FSU \cite{Todd2005} and NL3* \cite{Lalazissis2009} parameter sets, respectively. The referenced experimental data are taken from the Refs. \cite{Angeli2013,Audi2012}.

Table.~\ref{table2} The nuclear quadrupole moments $Q$ (units in fm$^2$) extracted from the electron scattering in this paper.  The experimental data is taken from the Ref. \cite{Stone2005}.

Table.~\ref{table3} The nuclear magnetic moments $\mu_\mathrm{sp}$ (units in nuclear magneton $\mu_\mathrm{N}$) calculated by Eq. (\ref{Mmoments}) with the FSU and NL3* parameter sets.  The experimental magnetic moments $\mu_\mathrm{exp}$ are taken from the Refs. \cite{Stone2005,Raghavan1989}.

Table.~\ref{table4} The spectroscopic factors for the magnetic form factors of $^{27}$Al and $^{39}$K. $\alpha_1$ is fixed to be $\alpha_1\equiv\mu_\mathrm{exp}/\mu_\mathrm{sp}$ and other $\alpha_L$s are obtained by the best fitting to the experimental data.

\newpage

\begin{table}[htb]
\centering \caption{Binding energies per nucleon $B/A$ (MeV) and charge radii $R_C$ (fm) of nuclei $^{14}$N, $^{27}$Al and $^{39}$K, calculated under the RMF theory  with FSU \cite{Todd2005} and NL3* \cite{Lalazissis2009} parameter sets, respectively. The referenced experimental data are taken from the Refs. \cite{Angeli2013,Audi2012}.  }\label{table1}\vspace{0.5cm}
\begin{tabular*}{8.5cm}{*{8}{c @{\extracolsep\fill}}}
\hline\hline
&\multicolumn{3}{c}{$B/A$ (MeV)}&&\multicolumn{3}{c}{ $R_C$ (fm)}\\
\cline{2-4}\cline{6-8}
Nucleus&$^{14}$N&$^{27}$Al&$^{39}$K&&$^{14}$N&$^{27}$Al&$^{39}$K\\
\hline
FSU&7.38&8.14 &8.49 &&2.62&3.02 &3.40 \\
NL3*&7.50 &8.10 &8.53 &&2.64 &3.00 &3.43 \\
Expt.&7.48&8.33&8.56&&2.56&3.06&3.44\\	
\hline\hline
\end{tabular*}
\end{table}

\newpage

\begin{table}[htb]
\centering \caption{The nuclear quadrupole moments $Q$ (units in fm$^2$) extracted from the electron scattering in this paper.  The experimental data is taken from the Ref. \cite{Stone2005}.  }\label{table2}\vspace{0.5cm}
\begin{tabular*}{8.6cm}{*{4}{c @{\extracolsep\fill}}}
  \hline \hline
  $Q$ & $^{14}$N & $^{27}$Al & $^{39}$K \\
  \hline
  Extracted &3.2&8.3&3.9\\
  Experimental &2.0&14.0&4.9\\
\hline\hline
 \end{tabular*}
\end{table}

\newpage

\begin{table}[htb]
\centering \caption{The nuclear magnetic moments $\mu_\mathrm{sp}$ (units in nuclear magneton $\mu_\mathrm{N}$) calculated by Eq. (\ref{Mmoments}) with the FSU and NL3* parameter sets.  The experimental magnetic moments $\mu_\mathrm{exp}$ are taken from the Refs. \cite{Stone2005,Raghavan1989}. }\label{table3}\vspace{0.5cm}
\begin{tabular*}{8.6cm}{*{4}{c @{\extracolsep\fill}}}
  \hline \hline
  $\mu$ & FSU & NL3* & Expt. \\
  \hline
  $^{27}$Al &5.36&5.53&3.64\\
  $^{39}$K &0.75&0.91&0.39\\
\hline\hline
 \end{tabular*}
\end{table}

\newpage

\begin{table}[htb]
\centering \caption{The spectroscopic factors for the magnetic form factors of $^{27}$Al and $^{39}$K. $\alpha_1$ is fixed to be $\alpha_1\equiv\mu_\mathrm{exp}/\mu_\mathrm{sp}$ and other $\alpha_L$s are obtained by the best fitting to the experimental data.}\label{table4}\vspace{0.5cm}
\begin{tabular*}{8.5cm}{*{7}{c @{\extracolsep\fill}}}
\hline\hline
Nucleus&\multicolumn{3}{c}{$^{27}$Al}&&\multicolumn{2}{c}{ $^{39}$K}\\
\cline{2-4}\cline{6-7}
&$\alpha_1$&$\alpha_3$&$\alpha_5$&&$\alpha_1$&$\alpha_3$\\
\hline
FSU&0.679&0.469 &0.630 &&0.526&0.584 \\
NL3*&0.658 &0.414 &0.573 &&0.430 &0.553\\
\hline\hline
\end{tabular*}
\end{table}

\newpage

\begin{figure}[htb]
\begin{center}
\includegraphics[width=12cm]{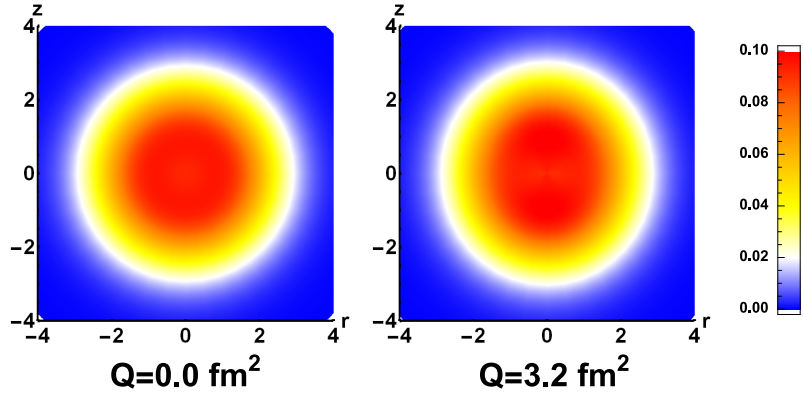}
\end{center}
\caption{Two-dimensional proton density distributions $\rho_p(\mathbf{r})$  of $^{14}$N in the $r-z$ plane of cylindrical coordinates, calculated by the Eq. (\ref{2pF_decom_pro_2}) with FSU parameter set. The corresponding quadrupole moments of two density distributions are $Q=0.0\ \mathrm{fm}^2$ and $Q=3.2\ \mathrm{fm}^2$, respectively.}\label{fig:density14N}
\end{figure}

Figure captions:

Fig.~\ref{fig:density14N} Two-dimensional proton density distributions $\rho_p(\mathbf{r})$  of $^{14}$N in the $r-z$ plane of cylindrical coordinates, calculated by the Eq. (\ref{2pF_decom_pro_2}) with FSU parameter set. The corresponding quadrupole moments of two density distributions are $Q=0.0\ \mathrm{fm}^2$ and $Q=3.2\ \mathrm{fm}^2$, respectively.

\begin{figure}[htb]
\begin{center}
\includegraphics[width=12cm]{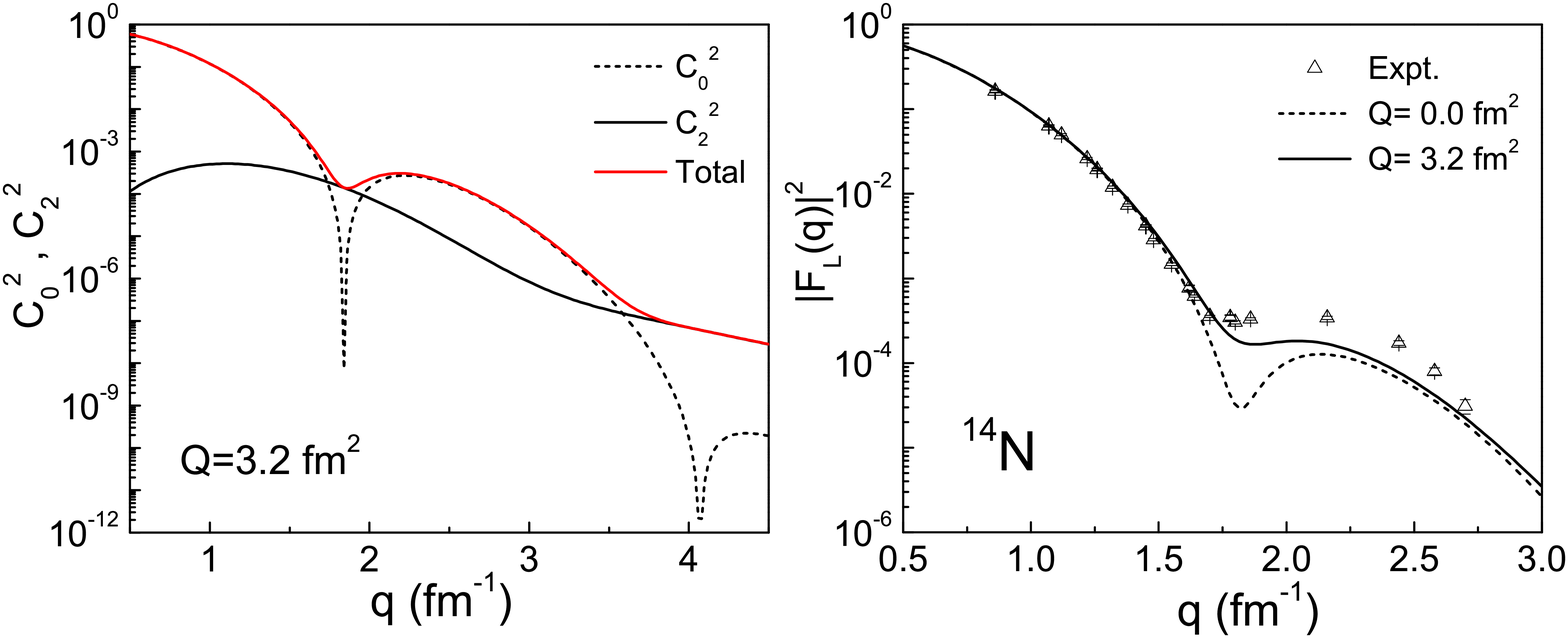}
\end{center}
\caption{Longitudinal form factors of $^{14}$N where the corresponding proton density distribution are calculated from the RMF theory with FSU parameter set. Left panel: $C_0$, $C_2$ and total form factor from PWBA method. Right panel: Nuclear longitudinal form factors for $Q=0.0\ \mathrm{fm}^2$ and $Q=3.2\ \mathrm{fm}^2$ where the Coulomb distortion corrections are taken into accounts. The Experimental data are taken from the Ref. \cite{Dally1970}.}\label{fig:N14FSU}
\end{figure}

Fig.~\ref{fig:N14FSU} Longitudinal form factors of $^{14}$N where the corresponding proton density distribution are calculated from the RMF theory with FSU parameter set. Left panel: $C_0$, $C_2$ and total form factor from PWBA method. Right panel: Nuclear longitudinal form factors for $Q=0.0\ \mathrm{fm}^2$ and $Q=3.2\ \mathrm{fm}^2$ where the Coulomb distortion corrections are taken into accounts. The Experimental data are taken from the Ref. \cite{Dally1970}.

\begin{figure}[htb]
\begin{center}
\includegraphics[width=12cm]{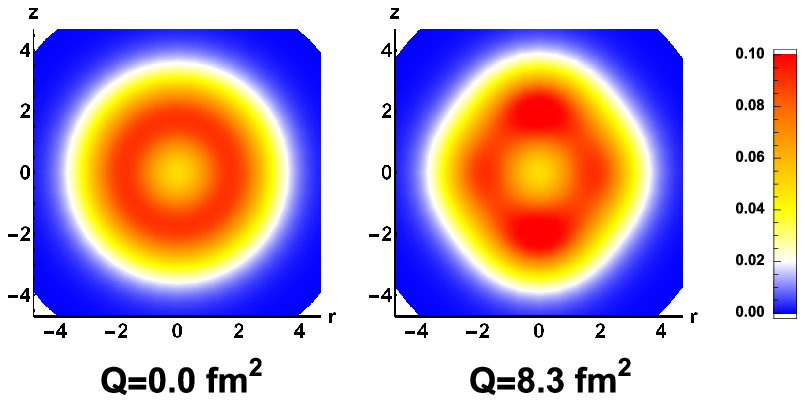}
\end{center}
\caption{Two-dimensional proton density distributions $\rho_p(\mathbf{r})$  of $^{27}$Al in the $r-z$ plane of cylindrical coordinates, calculated by the Eq. (\ref{2pF_decom_pro_2}) with FSU parameter set. The corresponding quadrupole moments are $Q=0.0\ \mathrm{fm}^2$ and $Q=8.3\ \mathrm{fm}^2$, respectively.}\label{fig:density27Al}
\end{figure}

Fig.~\ref{fig:density27Al} Two-dimensional proton density distributions $\rho_p(\mathbf{r})$  of $^{27}$Al in the $r-z$ plane of cylindrical coordinates, calculated by the Eq. (\ref{2pF_decom_pro_2}) with FSU parameter set. The corresponding quadrupole moments are $Q=0.0\ \mathrm{fm}^2$ and $Q=8.3\ \mathrm{fm}^2$, respectively.

\begin{figure}[htb]
\begin{center}
\includegraphics[width=12cm]{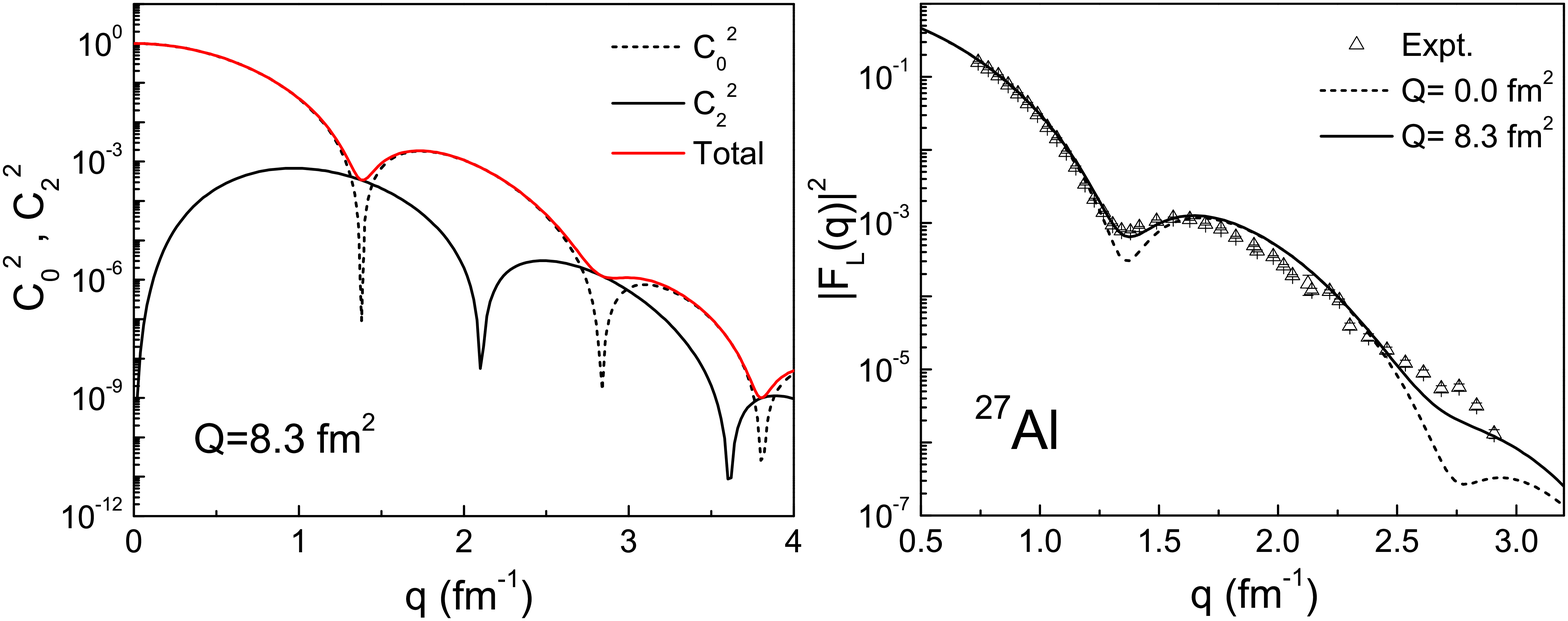}
\end{center}
\caption{Longitudinal form factors of $^{27}$Al where the corresponding proton density distribution are calculated from the RMF theory with FSU parameter set. Left panel: $C_0$, $C_2$ and total form factor from PWBA method. Right panel: Nuclear longitudinal form factors for $Q=0.0\ \mathrm{fm}^2$ and $Q=8.3\ \mathrm{fm}^2$ where the Coulomb distortion corrections are taken into accounts. The Experimental data are taken from the Ref. \cite{Yearian1974}.}\label{fig:Al27FSU}
\end{figure}

Fig.~\ref{fig:Al27FSU} Longitudinal form factors of $^{27}$Al where the corresponding proton density distribution are calculated from the RMF theory with FSU parameter set. Left panel: $C_0$, $C_2$ and total form factor from PWBA method. Right panel: Nuclear longitudinal form factors for $Q=0.0\ \mathrm{fm}^2$ and $Q=8.3\ \mathrm{fm}^2$ where the Coulomb distortion corrections are taken into accounts. The Experimental data are taken from the Ref. \cite{Yearian1974}.

\begin{figure}[htb]
\begin{center}
\includegraphics[width=12cm]{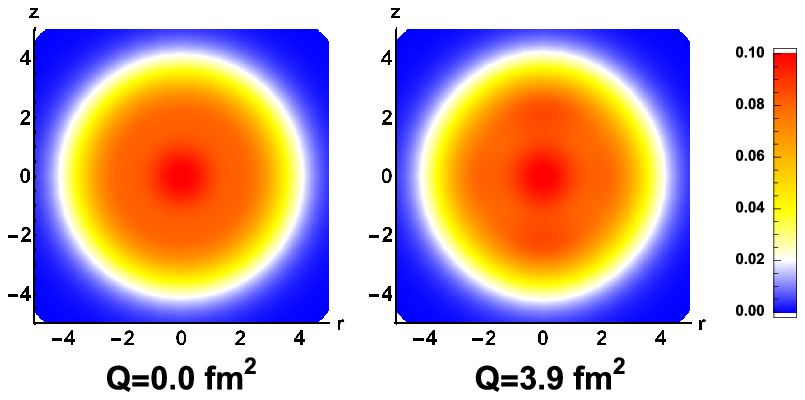}
\end{center}
\caption{Two-dimensional proton density distributions $\rho_p(\mathbf{r})$  of $^{39}$K in the $r-z$ plane of cylindrical coordinates, calculated by the Eq. (\ref{2pF_decom_pro_2}) with FSU parameter set. The corresponding quadrupole moments are $Q=0.0\ \mathrm{fm}^2$ and $Q=3.9\ \mathrm{fm}^2$, respectively.}\label{fig:density39K}
\end{figure}

Fig.~\ref{fig:density39K} Two-dimensional proton density distributions $\rho_p(\mathbf{r})$  of $^{39}$K in the $r-z$ plane of cylindrical coordinates, calculated by the Eq. (\ref{2pF_decom_pro_2}) with FSU parameter set. The corresponding quadrupole moments are $Q=0.0\ \mathrm{fm}^2$ and $Q=3.9\ \mathrm{fm}^2$, respectively.

\begin{figure}[htb]
\begin{center}
\includegraphics[width=12cm]{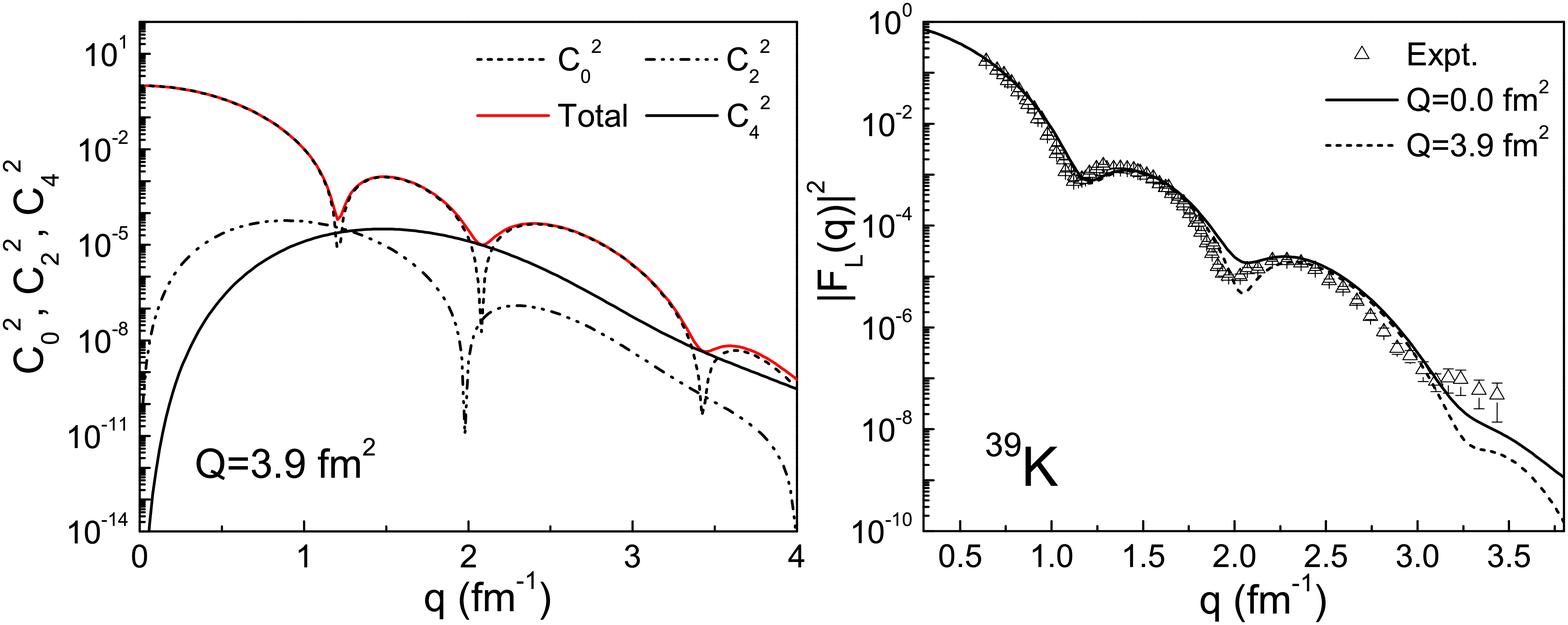}
\end{center}
\caption{Longitudinal form factors of $^{39}$K where the corresponding proton density distribution are calculated from the RMF theory with FSU parameter set. Left panel: $C_0$, $C_2$ and total form factor from PWBA method. Right panel: Nuclear longitudinal form factors for $Q=0.0\ \mathrm{fm}^2$ and $Q=3.9\ \mathrm{fm}^2$ where the Coulomb distortion corrections are taken into accounts. The Experimental data are taken from the Ref. \cite{Sinha1973}.}\label{fig:K39FSU}
\end{figure}

Fig.~\ref{fig:K39FSU} Longitudinal form factors of $^{39}$K where the corresponding proton density distribution are calculated from the RMF theory with FSU parameter set. Left panel: $C_0$, $C_2$ and total form factor from PWBA method. Right panel: Nuclear longitudinal form factors for $Q=0.0\ \mathrm{fm}^2$ and $Q=3.9\ \mathrm{fm}^2$ where the Coulomb distortion corrections are taken into accounts. The Experimental data are taken from the Ref. \cite{Sinha1973}.

\begin{figure}[htb]
\begin{center}
\includegraphics[width=12cm]{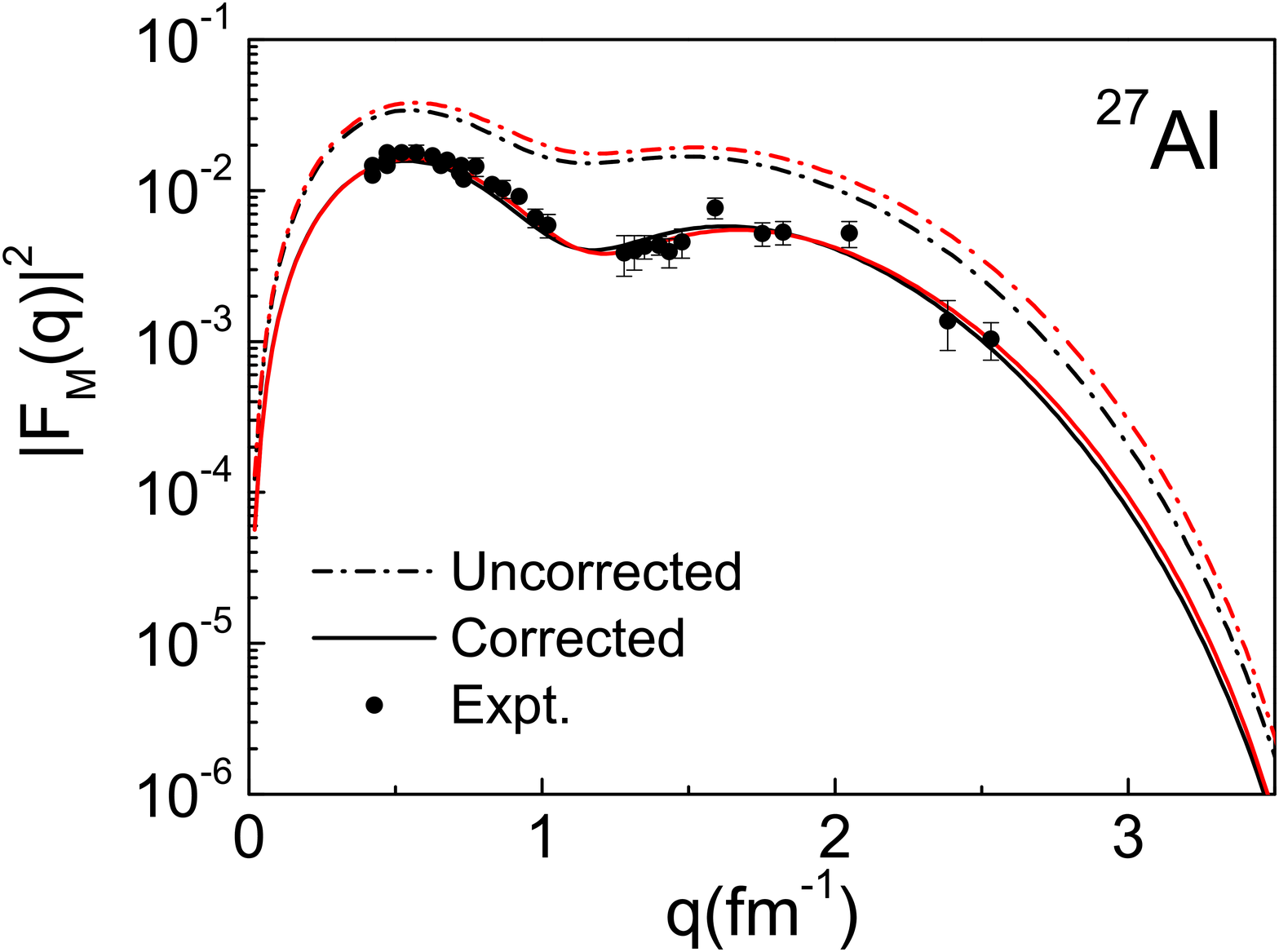}
\end{center}
\caption{The magnetic form factor of $^{27}$Al. The dash-dot lines represent the results calculated from Eq. (\ref{magform2}) without corrections. The solid lines represent the results calculated from Eq. (\ref{magform3}) including the spectroscopic factors. The black and red lines correspond to the calculations with FSU and NL3* parameter sets, respectively. The experimental data is taken from the Ref. \cite{Lapikas1973}.}\label{fig:Mag27Al}
\end{figure}

Fig.~\ref{fig:Mag27Al} The magnetic form factor of $^{27}$Al. The dash-dot lines represent the results calculated from Eq. (\ref{magform2}) without corrections. The solid lines represent the results calculated from Eq. (\ref{magform3}) including the spectroscopic factors. The black and red lines correspond to the calculations with FSU and NL3* parameter sets, respectively. The experimental data is taken from the Ref. \cite{Lapikas1973}.

\begin{figure}[htb]
\begin{center}
\includegraphics[width=12cm]{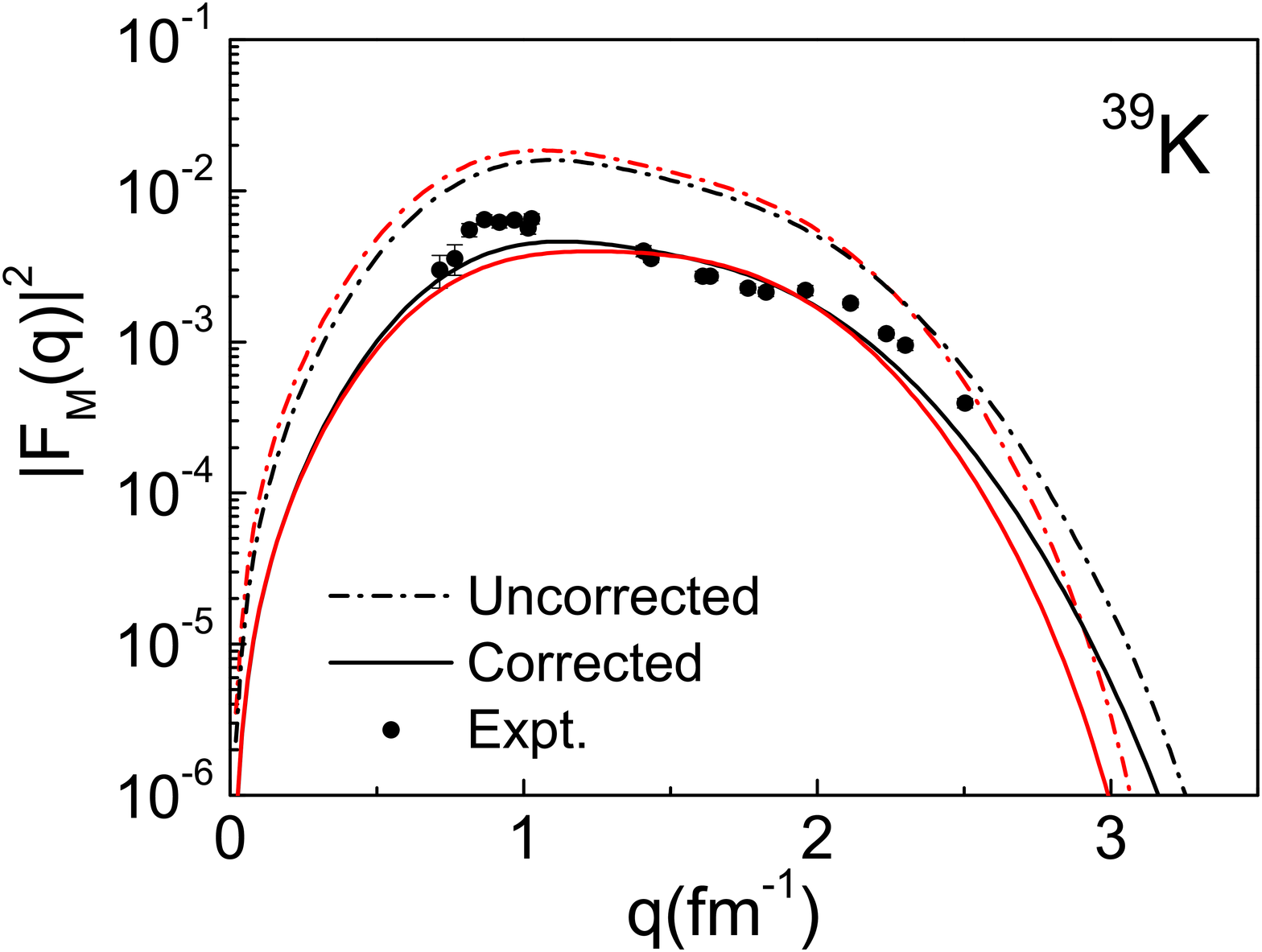}
\end{center}
\caption{The magnetic form factor of $^{39}$K. The dash-dot lines represent the results calculated from Eq. (\ref{magform2}) without corrections. The solid lines represent the results calculated from Eq. (\ref{magform3}) including the spectroscopic factors. The black and red lines correspond to the calculations with FSU and NL3* parameter sets, respectively. The experimental data is taken from the Ref. \cite{Donnelly1984}.}\label{fig:Mag39K}
\end{figure}

Fig.~\ref{fig:Mag39K} The magnetic form factor of $^{39}$K. The dash-dot lines represent the results calculated from Eq. (\ref{magform2}) without corrections. The solid lines represent the results calculated from Eq. (\ref{magform3}) including the spectroscopic factors. The black and red lines correspond to the calculations with FSU and NL3* parameter sets, respectively. The experimental data is taken from the Ref. \cite{Donnelly1984}.

\newpage

\newpage

\newpage

\newpage

\newpage

\newpage

\newpage

\newpage

\end{document}